# Strain effects in twisted spiral antimonene

Ding-Ming Huang*, Xu Wu, Kai Chang, Hao Hu, Ye-Liang Wang, H. Q. Xu*, Jian-Jun Zhang*

Dr. D.-M. Huang, Prof. K. Chang, Prof. H. Q. Xu, Prof. J.-J. Zhang
Beijing Academy of Quantum Information Sciences
Beijing 100193, China
E-mail: huangdm@baqis.ac.cn
E-mail: hqxu@pku.edu.cn
E-mail: jjzhang@iphy.ac.cn

Dr. D.-M. Huang, Prof. H. Q. Xu
Beijing Key Laboratory of Quantum Devices
Key Laboratory for the Physics and Chemistry of Nanodevices and School of Electronics
Peking University
Beijing 100871, China

Dr. D.-M. Huang, Prof. J.-J. Zhang
National Laboratory for Condensed Matter Physics and Institute of Physics
Chinese Academy of Sciences
Beijing 100190, China

Prof. X. Wu, Prof. Y.-L. Wang
MIIT Key Laboratory for Low-Dimensional Quantum Structure and Devices
School of Integrated Circuits and Electronics
Beijing Institute of Technology
Beijing 100081, China

Prof. H. Hu
Frontier Institute of Science and Technology
Xi'an Jiaotong University
Xi'an 710054, China

**Abstract:**

van der Waals (vdW) layered materials exhibit fruitful novel physical properties. The energy band of such materials depends strongly on their structures and a tremendous variation in their physical properties can be deduced from a tiny change in inter-layer spacing, twist angle, or in-plane strain. In this work, a kind of vdW layered material of spiral antimonene is constructed, and the strain effects in the material are studied. The spiral antimonene is grown on a germanium (Ge) substrate and is induced by a helical dislocation penetrating through few-atomic-layers of antimonene (β-phase). The as-grown spiral is intrinsically strained and the lattice distortion is found to be pinned around the dislocation. Both spontaneous inter-layer twist and in-plane anisotropic strain are observed in scanning tunneling microscope (STM) measurements. The strain in the spiral antimonene can be significantly modified by STM tip interaction, leading to a variation in the surface electronic density of states (DOS) and a large modification in the work function of up to a few hundreds of milli-electron-volts (meV). Those strain effects are expected to have potential applications in building up novel piezoelectric devices.

Two-dimensional (2D) van der Waals (vdW) layered materials are widely studied in the past two decades. [1~8,10~18] These materials are vdW stackings of atomic layers of 2D covalently bonded materials and their electronic property significantly relies on the inter-layer coupling. A certain setting of the inter-layer coupling or stacking order in such materials may emerge exotic physical properties. Twisted graphene stacking layers are of a well-known case, which exhibits superconductivity,[1] quantized anomalous Hall states [2] or electronic nematicity [3] under different stacking orders. A tiny change in the inter-layer coupling can lead to a significant change in the band structure, such as that the flat band can be induced by the inter-layer compression. [4,5] The in-plane strain, either a global strain or a local strain induced by STM tip interaction, in the stacking layers also has a large impact on the band structure [6~8] and thus electronic properties. As an example, the shift of van-Hove singularities in graphene stacking layers induced by a local strain has been observed.[7] Although the strain effects are widely studied on numerous 2D vdW layers systems, relevant research on spiral structures [9] is rarely reported. These spirals are of a kind of vdW layered materials with a helical dislocation penetrated through them. There may exhibit novel strain effects, as the interaction between layers is affected by the helical dislocation.

Recently, antimonene (β-phase) stacking layers have attracted great research interest. Antimonene is a buckled honeycomb 2D material of the antimony (Sb). It is expected to have applications for 2D photoelectric devices, owing to the semiconducting feature of the free-standing antimonene. [10] Exotic properties were observed in different antimonene stacking layers, such as semiconductor-semimetal transitions, [10,11] topological surface states, [12,13] and flat bands. [14,15] Few-layers antimonene is also predicted to exhibit fruitful strain effects, such as a band gap modulation induced by inter-layer compression, [11] and the quantum spin Hall state induced by in-plane strain. [16~18] Thus, spiral antimonene is a proper candidate for studying the strain effects on spirals.

Here, we report the epitaxial growth of spiral antimonene on Ge substrate by molecular beam epitaxy (MBE). Inter-layer twist forms in the spiral, as it is revealed by the moiré pattern. Quasi-particle interference (QPI) measurements show that the as-grown spiral is anisotropically strained. The strain in the spiral can also be manipulated by STM tip, as a tip induced large variation in the surface electronic density of states (DOS) and the work function are found. These strain-dependent properties are expected to have potential applications in constructing novel piezoelectric devices.

The spiral antimonene is grown on a Ge (111) substrate. Before Sb deposition, a monolayer of arsenic (As) is deposited on the Ge substrate at 920 K to form an unreconstructed As-terminated (1×1) surface. Fig. 1a is the schematics showing the top and side view of antimonene formed on As/Ge surface. The left part in the top panel of Fig. 1a is uncovered with antimonene to make the bottom As and Ge atoms clearly visible. Each As atom has five valence electrons and three of them are bonded to Ge in the second layer, leaving two electrons forming a lone-pair on the surface.[19] Thus, a surface with no dangling bonds is obtained, which is an ideal platform for van der Waals epitaxy (vdWE) of antimonene. The surface reconstruction during As covering is monitored by in-situ reflection high-energy electron diffraction (RHEED), which shows the transition from the "×2" pattern of Ge (111) c (2×8) reconstruction to "×1" unreconstructed pattern (Fig. 1b). The distribution of atomic steps can be controlled by As beam flux during growth. Bi-layer deep "step loops" are formed at a temperature of 920 K and an effective As flux pressure of $8.0×10^{-6}$ mbar. An STM image of "step loops" is shown in Fig. 1c. The (1×1) surface is confirmed by the atomic resolution STM image shown in the inset of Fig. 1c. Such bi-layer deep and dangling-bond free "step loops" are used to grow spiral antimonene.

Sb is deposited in two steps. At first, 0.05 ML of Sb is deposited at 280 K to form Sb clusters. The sample is then heated to 400 K and exposed to Sb beam flux for 10 min under an effective Sb pressure of $3.0×10^{-7}$ mbar. Antimonene spirals are observed on the As/Ge surface as helical dislocation is formed inside "step loops". The formation of dislocation is discussed in Section I of Supporting Information. The STM topographic images after Sb deposition are shown in Fig. 1d and 1e with different scales, where the helical dislocation is indicated by an arrow. The left-top inset of Fig. 1e shows the 3D view of the spiral. The measured step height on the spiral is 4.0 Å, which is equal to one atomic layer of the antimonene. The average thickness of spirals grown in this work is 10 atomic layers. The density of helical dislocations is about 50 $\mu m^{-2}$, which is similar to the density of surface "step loops". The RHEED diffraction pattern after Sb deposition is displayed in Fig. 1b. The clear (1×1) pattern indicates that all spirals have the same crystal orientation. The lattice constant of the antimonene is measured as 4.0 Å, indicating that it is under a large compressive strain (3 % smaller than the theoretical value 4.12 Å of free standing antimonene [20]). Such a large strain limits their lateral growth. The lateral growth continues only after the formation of helical dislocation, which releases the strain energy. This also explains why the lateral size of the spirals can reach to about 100 nm, while the diameter of antimonene flakes between spirals is less than 30 nm, as shown in inset of Fig. 1d.

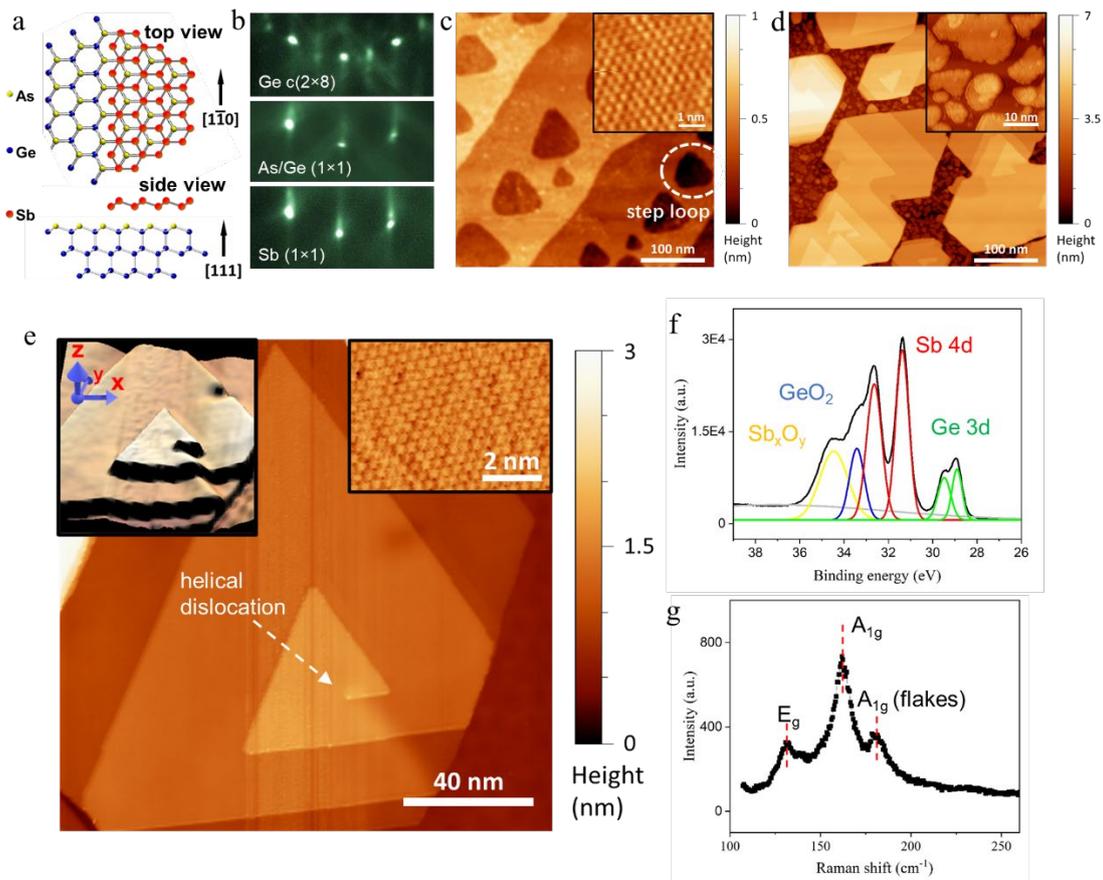

**Figure 1.** Spiral antimonene on a Ge (111) substrate. a) Schematic of the antimonene (partially covered) on an As/Ge surface. b) RHEED patterns of the Ge (111) c(2×8) surface, the As/Ge (1×1) surface and the spiral antimonene surface, respectively. Surface reconstruction is vanished once the substrate is fully covered with As. The "×1" feature of the RHEED pattern seen in the lowest panel of b) indicates that all antimonene spirals have the same crystal orientation. c) STM topographic image of the As/Ge surface with "step loops" (taken at 1.03 V and 65 pA). The inset shows an atomic resolution image of the top As atomic layer. d) STM topographic image of antimonene spirals (taken at 1.28 V and 95 pA). The inset shows antimonene flakes between spirals. e) STM image of a spiral (taken at -1.86 V and -150 pA), where a helical dislocation is marked by an arrow. Top-left inset: 3D view. Top-right inset: atomic resolution STM image of the antimonene buckled honeycomb lattice. f) XPS data of the spirals. Two main peaks (red fitted curves) correspond to Sb 4d orbits in antimonene, with binding energies of 31.4 eV and 32.6 eV, respectively. g) Raman spectrum of the spirals. Three peaks from left to right arise from the $E_g$ and $A_{1g}$ modes of the spirals and the $A_{1g}$ mode of the antimonene flakes, respectively. The Raman shift is accordant with β-phase antimonene.

The as-grown antimonene is characterized by in-situ STM, ex-situ X-ray photoelectron spectroscopy (XPS) and Raman spectrum. The atomic resolution STM measurements show a buckled honeycomb lattice (the inset of Fig. 1e). The XPS data is shown in Fig. 1f, where the two main peaks (red) seen at 31.4 eV and 32.6 eV are identified to be from Sb 4d orbits.[21] The peaks on the lower energy side of the main peaks are from Ge 3d orbits in the substrate (green), while the peaks on the higher energy side of the main peaks are from $GeO_2$ (blue) and $Sb_xO_y$ (yellow) due to oxidization in air.[22,23] Figure 1g shows the Raman spectrum of the spirals. The in-plane vibration mode $E_g$ and out-of-plane vibration mode $A_{1g}$ of antimonene are observed. The Raman peak at 179.7

cm$^{-1}$ is the A$_{1g}$ mode for single layer antimonene flakes. The peaks at 131.5 cm$^{-1}$ and 161.6 cm$^{-1}$ are E$_g$ and A$_{1g}$ modes of the spirals, which have a redshift of the vibration frequency due to the stacking of 10 layers.[24] Both the XPS and the Raman spectrum are measured with a light spot of a few micrometers in diameter. The spectra are average results over thousands of spirals.

As shown in Fig. 2a, we see a moiré pattern on the spiral antimonene, indicating the existence of an inter-layer twist. The moiré periods are from 7 nm to 10 nm in different spirals, corresponding to interlayer twisted angles of 2.3° to 3.3°. The difference in twisted angle may originate from the random size of "step loops" on the substrate. The moiré periods of some spirals show also a small deformation, as illustrated by the solid line in Fig. 2a. Such a distortion is attributed to local strain caused by helical dislocation.

In-plane strain is able to modify the elastic scattering of surface electrons, which could result in symmetry breaking in QPI. [25, 26] Spatial dI/dV maps are performed at low temperature of 10 K to acquire QPI signal on spirals. The STM topography of a mapping region next to a helical dislocation is shown in Fig. 2b. In order to intuitively demonstrate the anisotropic QPI signal, dI/dV maps taken at sample bias of +510 mV and +590 mV are selected to show (Figs. 2c and 2d, respectively). Both are performed under condition of 34 mV modulated amplitude and constant tunneling current of +335 pA. The insets show the 2D fast Fourier transform (FFT) of the dI/dV maps. The six-fold symmetrical points in FFT images are signals from the moiré pattern (marked in inset of Fig. 2c), which are invariant as the sample voltage increases. There exists a contour ring encircling the Brillouin zone center in the FFT images (dashed line in inset of Fig. 2d), which is attributed to an intra-band scattering in a conduction band, as the diameter of the contour ring expands when the sample voltage increases. An oval shape of the contour ring indicates that the electron scattering is biaxially anisotropic, which is different from the six-fold symmetry of intrinsic antimonene stacking layers. The QPI dispersions are obtained by integrating the linecuts in FFT of dI/dV maps, and the bias voltage of the dI/dV maps is ramped from +460 mV to +620 mV (Fig. 2e). The electron-pocket-like energy dependence of the scattering vector (dashed lines in Fig. 2e) indicates a band minimum at $465 \pm 10$ meV. The detailed data processing can be found in Section II of Supporting Information. The surface DOS in the QPI measuring region is characterized by dI/dV spectrum measurement (Fig. 2f). We speculate that the DOS spike at 465 meV is caused by the band minimum, which is accordant to the data of QPI. The QPI dispersion is clearly anisotropic in two Γ-K directions (Fig. 2e), and the ratio of electron effective mass in these directions is calculated as 1.6:1 (SI. II). Such anisotropy is a strong evidence of strain, which is pinned around the helical dislocation and spreads over about 100 nanometers in lateral directions.

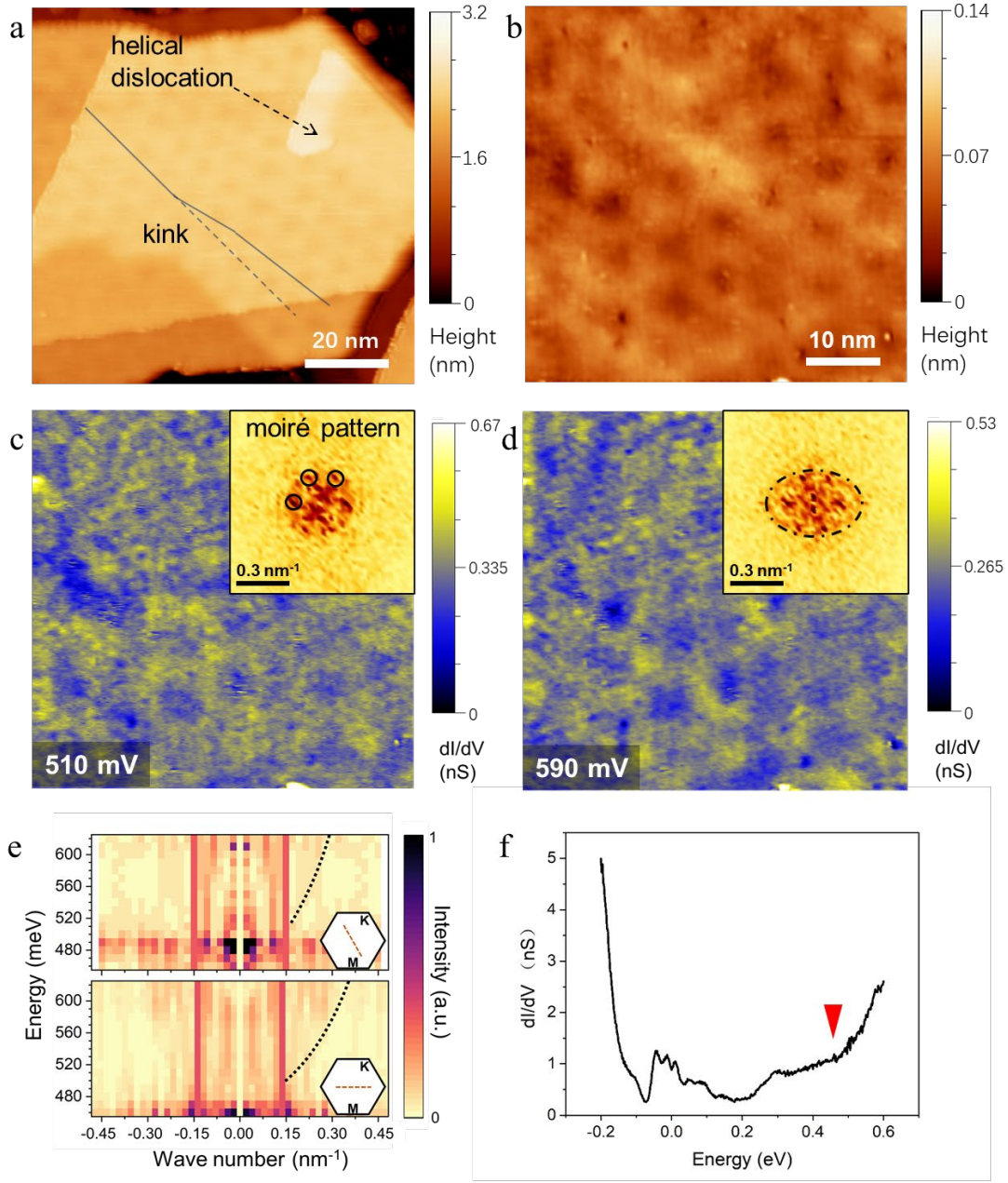

**Figure 2.** Strain in an antimonene spiral. a) Moiré lattice on the spiral. The solid line illustrates a kink on moiré pattern, corresponding to a lattice distortion near the helical dislocation. b) Topographic image of the QPI measuring region (taken at +510 mV and +335 pA). c,d) dI/dV spatial maps at sample voltages of +510 mV and +590 mV, respectively (with a 34 mV modulation and tunneling current of +335 pA). Insets: the corresponding FFT images. The oval contour ring in d) indicates the electron scattering is anisotropic. e) QPI dispersion. Dashed lines illustrate the energy level versus the wave number of intra-band scattering vectors in two Γ-K directions. f) dI/dV spectrum on the QPI measuring region (taken at +470 mV, +335 pA, and a 4 mV modulation). The red triangle marks the energy position of 465 meV at which the DOS turns to a quicker increase.

Previous studies reveal that a STM tip is able to manipulate the strain in graphene stacking layers, leading to variation in physical properties. [7,8] Strain in spiral antimonene can also be manipulated by a STM tip. We demonstrate that the surface DOS and the work function can be tuned by strain via performing "tip pressing" and "high voltage imaging" (HVI) on the spirals. Figure 3a schematically shows the processes of tip manipulation and property characterization. For "tip pressing", we directly reduce the tip height to the spiral surface. The strong repulsive force between the tip and the spiral induces distortion around the helical dislocation, such as a top layer twist of $1.0°\pm0.3°$ (Fig. 3b). Determination of the twist angle and the error analysis are detailed in Section III of Supporting Information. HVI is a conventional imaging process, with a high sample bias and a high tunneling current. Details of the tip pressing and the HVI are described in the Experimental Section. Characterizations of the DOS and the work function are performed next to the pressing position (with about 30 nm in lateral distance).

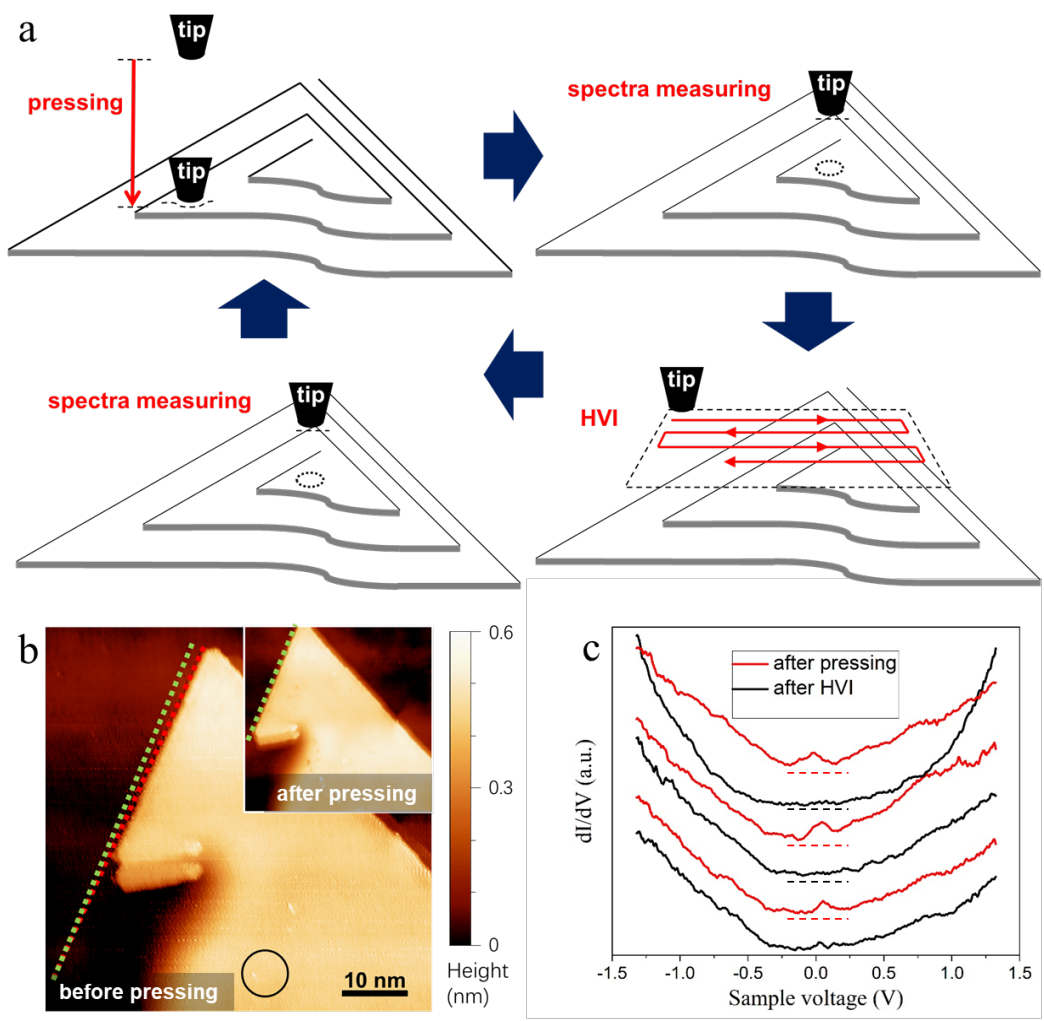

**Figure 3.** DOS modification induced by STM tip manipulating. a) Schematic diagram of the tip manipulating and the spectra measuring. b) The twist of the top layer after tip-pressing. Twist of $1.0°\pm0.3°$ is illustrated by the dashed line in STM image (taken at -1.3V and -102 pA). The pressing position is marked by circle. c) The DOS variation induced by tip manipulating. dI/dV spectra are measured on the spiral (-1.3 V, -102 pA, 15 mV modulation), during three cycles of tip pressing and HVI. A broad peak at 0 bias is observed after pressing (red) and disappeared after HVI (black).

Tip pressing and HVI are iteratively performed on a spiral. The surface DOS is characterized by dI/dV spectra. Six dI/dV spectra are displayed in Fig. 3c, which are measured in the order as illustrated in Fig. 3a. A broad peak near the Fermi-level is observed after tip pressing (red), the feature of the curves is similar to the spectrum shown in Fig. 2f. In contrast, the broad peak vanishes after HVI (black), and the feature is similar to the semi-metal few-layer antimonene in previous reports.[27] All of the dI/dV spectra are measured under the same tunneling condition of -1.3 V, -102 pA and a modulation amplitude of 15 mV.

The work function variation on a spiral is shown in Fig. 4a. The work function values are obtained by tunnel-junction-apparent-barrier-height in I-Z spectra. Detailed data processing and the error analysis are described in Section IV of Supporting Information. The work function decreases by hundreds of meV after tip pressing, and gets back after HVI. Each data point in Fig. 4a is an average result from 6 times of measurements, while the error bar is the root-mean-square (RMS) error. In addition, dZ/dV spectra are also measured to determine the work function variation on a spiral (Fig. 4b). The "Gundlach oscillation" peaks in dZ/dV spectrum correspond to the field resonance states in tunnel junction.[28] The energy shift of the peaks is equal to the difference of work function, as long as the electric field in tunnel junction is invariant in measurements.[29] The energy shifts induced by tip pressing and HVI in Fig. 4b are -0.26±0.01 eV and +0.1±0.01 eV, respectively. which further demonstrates the hundreds of meV of variation in the work function. The dZ/dV spectra are obtained by numerical differentiation of the Z-V spectra. More details of the work function measurements can be found in Section IV of Supporting Information.

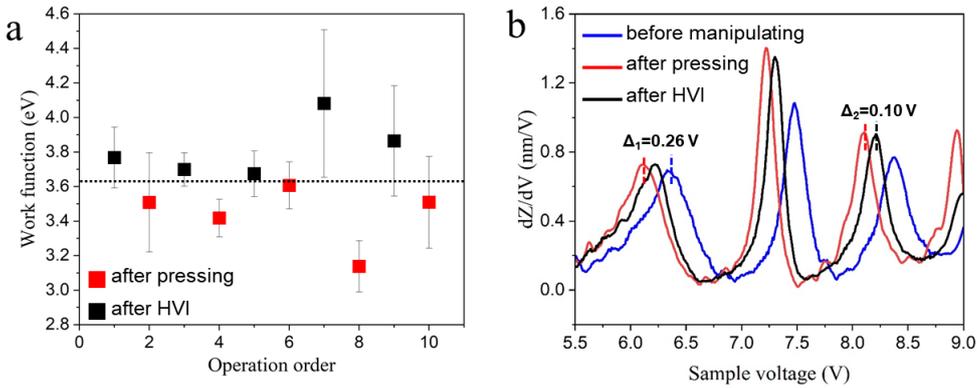

**Figure 4.** Work function variation induced by tip manipulating. a) Work function variation characterized by I-Z spectra. The X-coordinate is the operation order, tip manipulating of HVT and "tip pressing" are represented by odd and even numbers, respectively. Error bar is the RMS error from 6 times of measurements. b) Work function variation characterized by dZ/dV spectra. The energy shifts of the "Gundlach oscillation" peaks indicate work function variations of -0.26±0.01 eV and +0.1±0.01 eV, respectively.

The modifications of DOS and work function are considered to originate from the variation in strain. The strain induced by tip pressing can reside in a spiral, as the distortion can be pinned around the helical dislocation. In the other hand, the HVI generates a strong electric field between the tip and the spiral, and the spiral tends to release the pressing induced strain under such perturbation. The distortion in spiral can be classified into in-plane and inter-layer distortion, respectively. The in-plane distortion has been observed as the twist of the top layer (Fig. 3b). But there is lack of direct observation of the inter-layer distortion, as such sub-Angstrom change in height is too small to confirm by STM. We speculate that the inter-layer distortion has the main contribution to the variations of DOS and work function. A suppression of the inter-layer coupling, which is caused by an inter-layer distortion, may lead the top layer of the spiral to become nearly freestanding. The band structure of the nearly freestanding top layer should be similar to the semiconducting antimonene[10,11], that is why the electronic states near the Fermi-level is able to be manipulated by STM tip (Fig. 3c). The work function also significantly depends on the inter-layer coupling, a change of several hundred meV in work function can be achieved by layers stacking. [30~32] In the other hand, the work function of few-atomic-layers films is affected by quantum well states, which is significantly relied on the material thickness. [33] Although the tip induced inter-layer distortion is sub-Angstrom, the change in inter-layer coupling still may lead tremendous changes in the "effective thickness" and the work function. An indirect evidence of the tip induced inter-layer distortion is discussed in Section V of Supporting Information. In order to confirm that the variations of DOS and work function are caused by the STM tip manipulating, but not measurement error from local-states on moiré lattice, local electronic states and local work function are studied detailed for comparison in Section VI of Supporting Information.

In summary, we demonstrate the epitaxial growth of the spiral antimonene on Ge (111) substrate, and the strain related electronic properties of the spiral antimonene are investigated in nano scale. Intrinsic in-plane anisotropic strain is observed in the spiral, as the helical dislocation in the spiral is a natural pinning center for lattice distortion. The distortion induced by STM tip interaction is able to reside in the spiral, lending to a change of strain condition. The surface DOS and the work function of the spiral antimonene are relied on the strain, which can be repeatedly manipulated by STM tip interaction. Such strain-dependent modifications of DOS and work function are expected to have potential applications in novel piezoelectric devices or contact technic of nanodevices. The spiral also provides an ideal platform to study the strain effects around the lattice dislocation.

## Experimental section

*Growth*: The MBE is interconnected by two growth chambers, one for Ge deposition and the other for As, Sb deposition. At first, a Ge (111) substrate was heated at 890 K for 6 min in the Ge-growth chamber to remove the native oxidation. Then, the temperature was ramped down to 590 K for deposition of a 50 nm Ge buffer layer, and the growth rate was 0.5 Å/s. A flat Ge surface was achieved after an annealing process at 690 K for 5 min. The atomic steps on such surface were arranged in parallel array. After the buffer layer, the sample was transported to the As-Sb-growth chamber, and the background pressure was below $3.4\times10^{-9}$ mbar during the transport. The sample was exposed to an $As_4$ molecule flux with an effective pressure of $8.0\times10^{-6}$ mbar (measured by in-situ beam flux monitor), once it was loaded in the As-Sb-growth chamber. The $As_4$ flux was evaporated from an elementary cracker source. Then, the substrate temperature was rapidly ramped up to 920 K and held for 10 min under the $As_4$ flux. The Ge (111) surface was passivated by a single atomic layer of As atoms under such condition, and the "step loops" were formed at the same time. After the As passivation, we shut down the $As_4$ flux and ramped the temperature down to 280 K, in preparation for the followed Sb deposition. Sb was evaporated from an elementary cracker source. The Sb deposition was consisted of two steps: 0.05ML deposition of the nucleating layer at 280 K, followed by the growth of spiral antimonene at 400 K. The effective pressure of the Sb flux was controlled at $3.0\times10^{-7}$ mbar to achieve the growth of spirals.

*STM experiments*: The STM chamber is connected to the MBE chamber in vacuum. All of the STM measurements were performed under the low temperature of 10 K, and the temperature was held by a cryostat. The STM tip was a tungsten (W) tip, which was obtained by electrochemical corrosion. The tip was treated and spectroscopically characterized on the Au (111) surface before experiments, to acquire a standard tip for spectroscopic measurements. The dZ/dV spectra were obtained by numerical differentiation of the Z-V spectra. The Z-V spectra were measured by ramping the sample bias, in the condition of the tunneling current was fixed. The tip interaction of "tip pressing" were performed near the helical dislocation of spirals. Before tip pressing, the sample bias and tunneling current were set to -1.30 V and -102 pA, respectively. The feed-back of the tunneling current was turned off when the tip pressing was performed. Then, we directly reduced the tip height from the original tip position, and the descent of the tip was at least 2.5 nm for operations in this work. The sample bias was fixed during the pressing process. After the tip pressing, the tip was retracted and then reengaged for followed measurements. High voltage imaging (HVI) were performed as conventional constant current imaging process. The sample bias and tunneling current were set to +4 V and +1 nA, respectively. The number of lines per frame for a 60×60 nm$^2$ imaging region was at least 64 in the HVI process. All STM images in this paper were processed by the software named WSxM.[34]

## Supporting Information
Supporting Information is available from the Wiley Online Library or from the author.

## Acknowledgements
This work was supported by Innovation Program for Quantum Science and Technology (No. 2021ZD0302300), the NSFC (No. 92165207, No. 92165208, and No. 11874071), the Strategic Priority Research Program of CAS (Grant No. XDB30000000).

## Conflict of Interest
The authors declare no conflict of interest.

## Keywords
Antimonene, helical dislocation, spiral, strain effect, vdW layered material

# Supporting Information I:

The growth mechanism in this work is different from the previous study of spiral growth on non-Euclidean substrate surface.[9] Although the "step loops" on the Ge substrate are non-Euclidean structures, the inter-layer twist derived from this geometry is calculated as less than 1°, which is quite smaller than the value measured from the moiré pattern on spirals of 3°. The mechanism of step edge induced spiral growth [35] is also failed in this work. Antimonene is repelled to form helical dislocation, if there are only long parallel atomic steps present on the substrate (Fig. SI. 1a).

We speculate that the formation of helical dislocation is induced by strain accumulation inside the "step loop". The theoretical value of in-plane lattice constant of antimonene is 4.12 Å, which is about 3% larger than the Ge (111) surface. When the first layer antimonene grows and fills the "step loop", the lattice mismatch induced strain in antimonene neither relax at the edge due to constrain of "step loop" nor by forming undulation since we didn't observe it in the experiment, which is probably due to the relative strong interaction between antimonene and As/Ge(111) surface. Such large, accumulated compressive strain could induce novel structure formation. A possible way for the strain relaxation could be vacancy aggregation.[36] The Sb vacancies are tended to assemble into line structure under such large compressive strain and form a crack, as showed in Fig. SI. 1b. Further growth of the antimonene may lead to the step flow growth along one edge of the crack and formation of the spiral structure. The schematic diagram of spiral growth is illustrated in Fig. SI. 1c-g. Indeed, we found in the experiments that the density of the spiral is comparable to that of the "step loop", indicating that the spiral structure is induced by the surface "step loop". When we grow Sb on a flat As/Ge(111) surface (only low density intrinsic parallel atomic steps are presented on the substrate), we found that antimonene is tend to forming multilayers island but not spiral (Fig. SI. 1a), it is probably due to the strain relaxation induced by the free edge of the 2D antimonene island, which also support our proposal, that the growth of the spiral structure is induced by the strain accumulation of first layer antimonene in the "step loop".

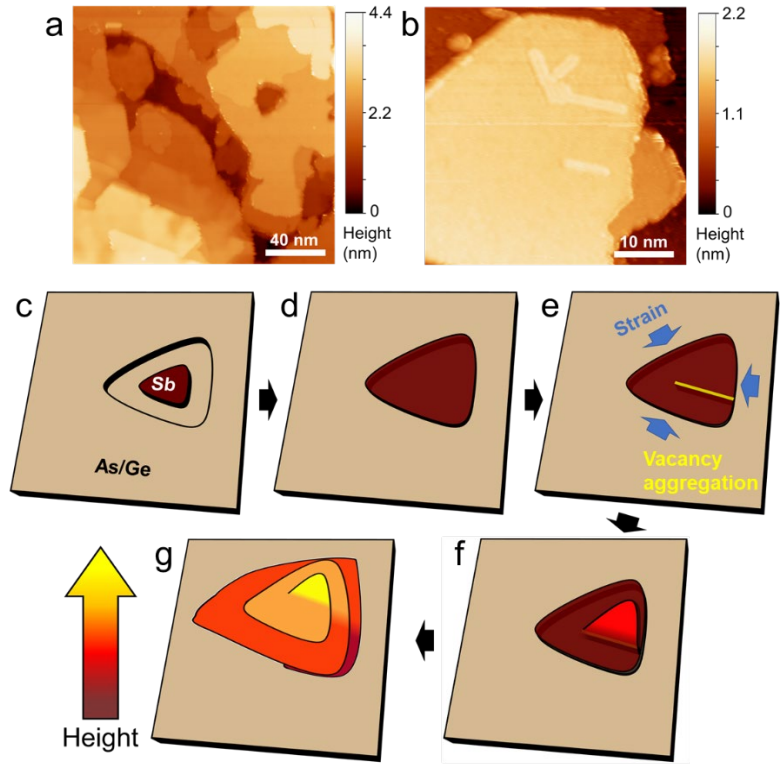

**Figure SI. 1.** Growth mechanism of the spiral antimonene. a) Multilayers islands of antimonene is formed when the substrate has no "step loop". b) Vacancy lines on antimonene. c~g) Schematic diagram of spiral growth. The compressive strain is accumulated as the antimonene flake expanding. Free edge relaxation is constrained by the "step loop", and vacancy lines formed. Helical dislocation is then formed on the vacancy line.

## Supporting Information II

The Fig. 2e. in the paper main text is obtained by integrating the linecut profiles from the FFT images of QPI dI/dV maps. The profiles are recalibrated by normalizing the intensity of moiré points.

The energy dependence of the intra-band scattering vector is fitted to quadratic equation:

$$E = A * q^2 + B$$

where, $E$ is the energy level. The wave number of scattering vector $q$ is determined as the maximal momenta of the QPI peaks in FFT profile, and the error bar is the scale of pixel in FFT images of 0.02 nm$^{-1}$. The fitting curves for two different Γ-K directions are illustrated in Fig. SI. 2a. The coefficient $B$ of these two fitting curves are determined as 470±5 meV and 459±9 meV, respectively. As we assume that the scattering is an intra-band scattering, the band minimum is equal to $B$. The band minimum is finally determined to be 465±10 meV.

The coefficient $A$ is related to electron effective mass $m_{eff}$, which has the relation of:

$$m_{eff} = \frac{\hbar^2}{8A}$$

The $m_{eff}$ in these two fitting curves are determined as 0.008 $m_e$ and 0.005 $m_e$, respectively. An asymmetry of the electron scattering behavior is clearly observed.

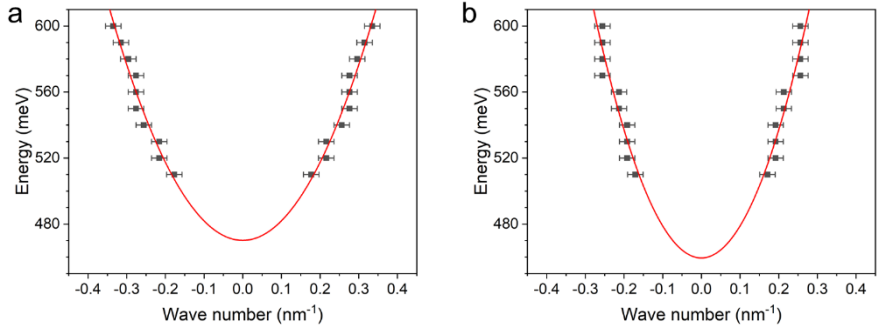

**Figure SI. 2.** Fitting curves of the energy dependence of the QPI scattering vector. a) and b) are fitting curves in two different Γ-K directions, respectively.

## Supporting Information III

We put the original data of STM images on Fig. 3b of the paper main text, for intuitively demonstrate the twist angle. It should be noted that the true twist angle is not accordant to the dash line in Fig. 3b, owing to the measurement error from double-tip effect and scanning shift. The double-tip effect lows down the contrast of the step edge, and the scanning shift arises a tiny deformation on the image. In order to eliminate the error from double-tip effect, we use the reciprocal signal of the step edge in FFT image to define the normal direction of the edge, and the low contrast details of the step edge can no longer affect the determination of the step direction. We assume that all of step edges on the spiral are zigzag edges. The error from scanning shift can be eliminated, by recalibrating the zigzag edges in STM and FFT images. The recalibrated topographic image of the spiral and the corresponding FFT image are showed in Fig. SI. 3a and b, respectively.

In order to determine the tip pressing induced twist, the variation of intensity distribution along a line section in FFT images are measured, which is showed in Fig. SI. 3c. The measuring section line is illustrated by a white dashed line in Fig. SI. 3b. The twist induced by tip pressing can be calculated by the shifts of the peaks in the FFT profiles (Fig. SI. 3c). The tip pressing induced twist is eventually determined to be $1.0° \pm 0.3°$.

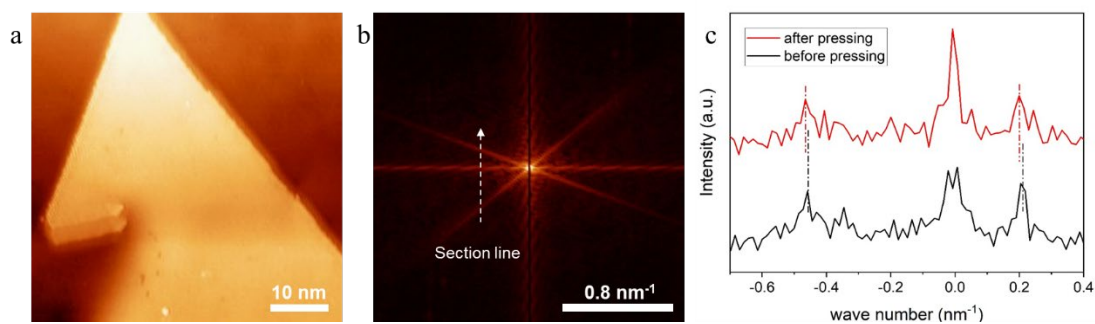

**Figure SI. 3.** Determination of the tip pressing induced twist. a) and b) Recalibrated STM image of the spiral and the FFT image, respectively. c) The section profiles along the white dashed line in b. Black and red curves are profiles before and after tip pressing, respectively.

## Supporting Information IV

The current in tunnel junction is dependent to the sample voltage $V$ and the barrier width $W$, by the relation of:

$$I \propto V\rho\, e^{-2\frac{\sqrt{2m_e\Phi}}{\hbar}W}$$

where, $\rho$ is the density of states of tunneling electron, $m_e$ is the mass of electron and $\Phi$ is the surface work function. The current $I$ is exponential decayed with increasing of $W$, once $V$ is fixed.

The work function $\Phi$ is characterized by I-Z spectra. The STM tip is grounded and the sample voltage is set at -1.3 V during measurements. The I-Z spectra is then fitted to the equation of:

$$I = A\, e^{-B*Z}$$

The work function $\Phi$ is obtained from the tunnel-junction-apparent-barrier-height, which is decided to the exponential decay factor of B:

$$\Phi = \frac{\hbar^2}{8m_e}B^2$$

All of the I-Z spectra in this work exhibit well exponential relation, and an example is shown in Fig. SI. 4a.

Each work function data point in Fig. 4a of the paper main text is averaged from 6 groups of I-Z spectra, and the RMS error of the data points are about few hundreds of meV. An apparent change in the work function of a spiral is presented only after a deep "tip pressing", that the tip height is reduced by at least 2 nm. The work function is constant during I-Z measurement, as the tip height is changed within 0.25 nm. The non-negligible RMS error is not a measurement error caused by tip induced work function variation, but a systemic error.

dZ/dV spectra are measured to determine the work function variation with higher precision. Once the electric potential in the tunnel junction is approximated by a triangular shape. The energy levels of the field resonance states in the triangular potential are given by:

$$E_n = \Phi + \alpha_n F^{2/3}$$

Where, $\Phi$ is the surface work function, $F$ is the electric field (which is a constant in a triangular potential), and $\alpha_n$ is the coefficient related to the quantum number $n$. [28,29]

The energy shift (ES) of the field resonance states between different materials is described by:

$$\Delta E_n = \Phi_1 - \Phi_2 + \alpha_n(F_1^{2/3} - F_2^{2/3})$$

In the condition of $F_1=F_2$, the ES of different resonance states are independent to the quantum number n, and the ES of all states are equal to the difference of the work function.

The "Gundlach oscillation peaks" in STM dZ/dV spectrum are caused by the field resonance states in tunnel junction. We can determine the work function variation by studying the ES of the "Gundlach oscillation peaks". Such method is generally applied to determine the work function in nano scale with precision of $\pm 10$ meV. [37~40]

The dZ/dV spectra in this paper are obtained by numerical differentiation of the Z-V spectra.

Three sorts of dZ/dV spectra on a spiral antimonene are shown in Fig. SI. 4b, which are measured along three times of STM tip interactions. The ES of all peaks in the black and red curves are equal to 0.08±0.01 eV, corresponding to a work function variation of 0.08 eV after the tip interaction. The ES of the peaks in the red and green curves are inconsistent for different field resonance states. The inconsistence is caused by the change of the electric field F in tunnel junction, which may originate from a considerable variation of work function too (~1 eV scale). The dZ/dV spectroscopy is no longer available to detect the work function variation in such condition.

In summary, I-Z spectrum is efficient to detect the work function variation in the energy scale of 0.1~1 eV, with the systemic error of few hundreds of meV. In the other hand, dZ/dV spectrum is efficient to detect a work function variation in the energy scale of ~0.1 eV, with a high precision of ±10 meV. We must use both of two methods to demonstrate the tip induced work function variation, as which is presented in the paper main text.

It is worth noting that the tip induced work function variation is widely reproducible, but the variation rules for different spirals are not consistent. The work function of a spiral may increase after tip pressing, which is totally opposite to the data in Fig. 4a. Such difference between spirals is comprehensible, as the helical dislocation is essentially a disorder in the lattice, and each spiral has its own unique stacking order and original strain condition.

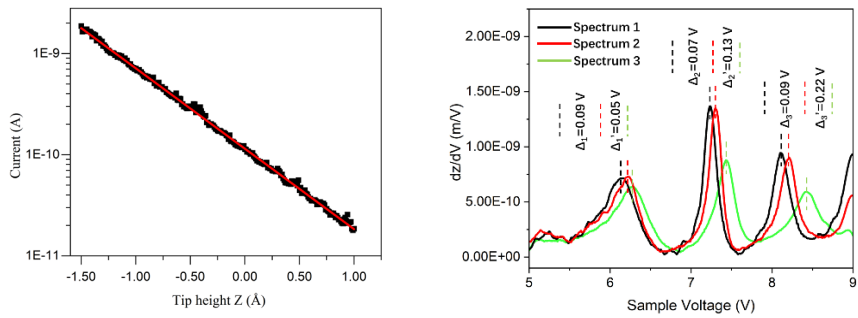

**Figure SI. 4.** Spectroscopy for work function determination. a) A representative I-Z spectrum in the logarithmic coordinates. b) The energy shift of the Gundlach oscillation peaks in three representative dZ/dV spectra.

## Supporting Information V

In order to demonstrate the effect of inter-layer coupling to the work function variation, a "dual-helical spiral" is studied. The STM morphology of the "dual-helical spiral" is shown in Fig. SI. 5a, there are two helical dislocations in the spiral. The Burgers vectors of these two dislocations are opposite to each other. A "single-helical spiral" is free to twist under external interaction, and the helical dislocation is the twist axis. But the "dual-helical spiral" is unable to twist, because the twists around these two dislocations are inter-locked by each other (Fig. SI. 5b). As a consequence, in-plane distortion is relatively hampered on the "dual-helical spiral". The tip induced work function variation of a "dual-helical spiral" is shown in Fig. SI. 5c. The "tip pressing" is obviously able to change the work function, but the HVI is almost failed.

Here we propose a hypothesis to understand the above-mentioned phenomenon: The distortion induced by "tip pressing" consist of two parts, in-plane and inter-layer distortion, respectively. Both of them can be resided in a "single-helical spiral", especially the in-plane distortion is correlated to the inter-layer twist. However, only the distortion of inter-layer spacing can be resided in a "dual-helical spiral", as the twist is prohibited. The "single-helical spiral" is not as stable as the "dual-helical spiral" after STM tip pressing, as a result of the combination of distortions. The tip induced force in HVI is quite weaker than that in "tip pressing". The interaction of HVI is strong enough to relax the distortion in "single-helical spiral", but it is unable to relax the more stable distortion in "dual-helical spiral". That is why the HVI is failed to change the work function of the "dual-helical spiral" (Fig. SI. 5c).

Under such hypothesis, the work function variation caused by "tip pressing" in Fig. SI. 5c is originated from the distortion of inter-layer spacing.

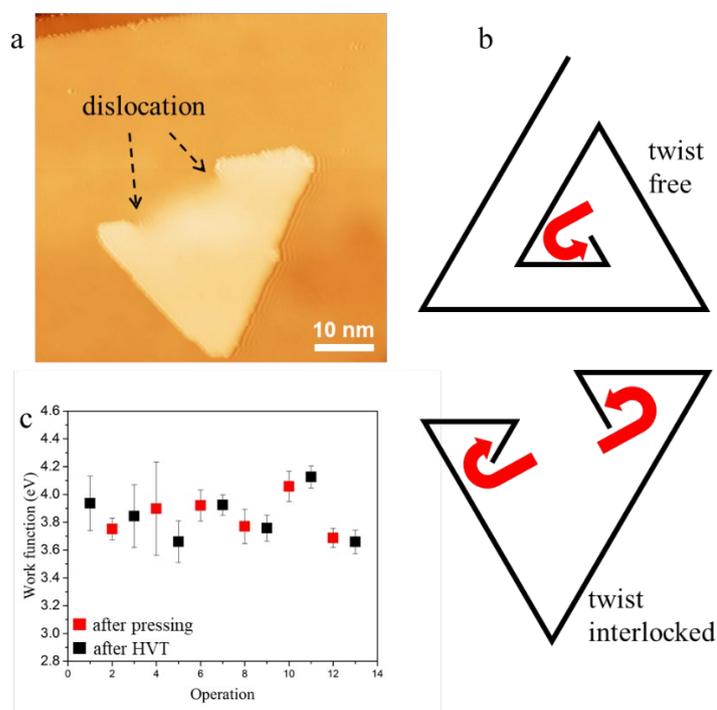

**Figure SI. 5.** Dual-helical spiral. a) STM topographic image of a "dual-helical spiral". b) Schematic diagram of the twist in "single-helical spiral" and "dual-helical spiral". c) Variation of the work function in a "dual-helical spiral".

## Supporting Information VI

In order to verify that the DOS variation in the paper main text is caused by distortion, but not a measurement error from the local-electronic-states on moiré superlattice, the dI/dV line map on the moiré pattern is studied. The line map is shown in Fig. SI. 6a and the measuring line is illustrated by dashed line in Fig. SI. 6c. We can clearly observe that only a few states near the Fermi-level are location-relevant (marked by red triangle in Fig. SI. 6a). The broad peak between 0 mV and 200 mV is almost independent to the location. We speculate that the local-electronic-states are dependent to the local stacking rule on the moiré pattern,[11] and the broad peak in the line map is caused by a global surface state. The local work functions on the moiré pattern are also studied, and the dZ/dV spectra are shown in Fig. SI. 6b. The measuring positions are marked by dots in Fig. SI. 6c. The energy shift of the "Gundlach oscillation peaks" is within 30 meV, indicating that the location-relevant work function variation is smaller than 30 meV. In summary, the tip induced variations of DOS and work function are indeed structure-distortion-relevant, and are independent to the local-states on moiré superlattice.

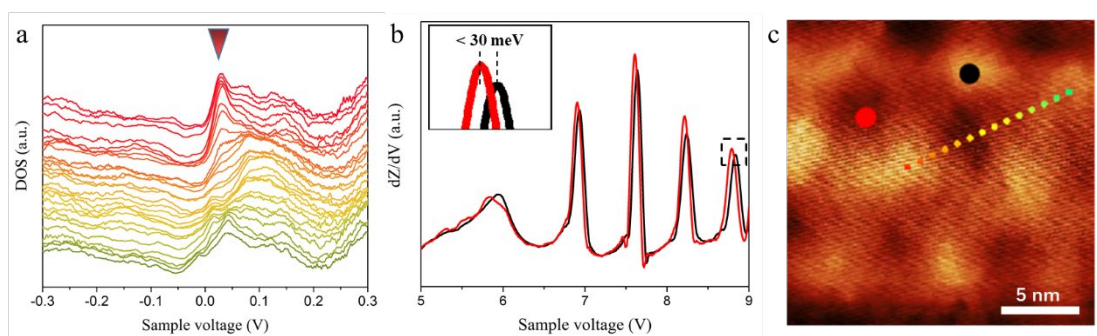

**Figure SI. 6.** Local-electronic states and local work functions on the moiré pattern. a) dI/dV line map on the moiré pattern. b) dZ/dV spectra on different positions. c) STM image of the moiré pattern. Measuring positions of the dI/dV line map and the dZ/dV spectra are marked by dashed line and dots, respectively.

Reference：